\documentclass[english,prd,nofootinbib,preprintnumbers,twocolumn]{revtex4}

\usepackage[latin1]{inputenc}
\usepackage{lmodern}
\usepackage{amsmath}
\usepackage{amsfonts}
\usepackage{appendix}
\usepackage{amssymb}
\usepackage{epsfig}
\usepackage{graphics,psfrag,rotating}
\usepackage{graphicx}
\usepackage{dcolumn}
\usepackage{bm}
\usepackage{epstopdf}
\usepackage{color}
\usepackage[usenames,dvipsnames,svgnames]{xcolor}

\usepackage[colorlinks=true,linktoc=all]{hyperref}
\usepackage{hyperref}
\usepackage[T1]{fontenc}
\usepackage{multirow}
\usepackage{float}
\usepackage{cancel}
\usepackage{hyperref}
\usepackage{adjustbox}
\usepackage{subfigure}

\usepackage{enumitem}

\def\3nab{\tilde{\nabla}}

\def\be {\begin{equation}}
\def\ee {\end{equation}}
\def\ba {\begin{align}}
\def\ea {\end{align}}

\def\bc {\begin{center}}
\def\ec {\end{center}}
\def\case#1/#2{\frac{#1}{#2}}

\newcommand{\bver}{\begin{verbatim}}
\newcommand{\ever}{\end{verbatim}}

\newcommand{\bea}{\begin{eqnarray}}
\newcommand{\eea}{\end{eqnarray}}
\newcommand{\beaa}{\begin{eqnarray*}}
\newcommand{\eeaa}{\end{eqnarray*}}

\newcommand{\lsim}   {\mathrel{\mathop{\kern 0pt \rlap
  {\raise.2ex\hbox{$<$}}}
  \lower.9ex\hbox{\kern-.190em $\sim$}}}
\newcommand{\gsim}   {\mathrel{\mathop{\kern 0pt \rlap
  {\raise.2ex\hbox{$>$}}}
  \lower.9ex\hbox{\kern-.190em $\sim$}}}
\def\case#1/#2{\textstyle\frac{#1}{#2}}

\begin{document}

\title{$\Lambda$CDM from broken diffeomorphisms}

\author{Javier de Cruz P\'erez}
\email{jdecruz@uco.es}
\affiliation{Departamento de F\'{\i}sica, Universidad de C\'ordoba,
Campus Universitario de Rabanales, Ctra. N-IV km, 396, E-14071, C\'ordoba, Spain}

\author{Antonio L.\ Maroto}
\email{maroto@ucm.es}
\affiliation{Departamento de F\'{\i}sica Te\'orica and\\ Instituto de F\'{\i}sica de Part\'{\i}culas y del Cosmos (IPARCOS-UCM), Universidad Complutense de Madrid, 28040 
Madrid, Spain}



\date{\today}

\begin{abstract} 
We present a simple field theory model with reduced invariance under diffeomorphisms whose
energy-momentum tensor is identical to the sum of  pressureless irrotational matter and a
cosmological constant. The model action is built from a single scalar field with a canonical kinetic term without any potential or Lagrange multiplier terms. The coupling to gravity is realized through a particular transverse diffeomorphism invariant volume element. The corresponding sound speed is exactly zero in any background geometry and the model is dynamically identical to $\Lambda$CDM. By restoring the full diffeomorphism invariance through the introduction of Stueckelberg-like fields, we obtain an equivalent local scalar-vector theory.   
\end{abstract} 
\preprint{IPARCOS-UCM-25-022}
\maketitle 
\section{Introduction}

Diffeomorphism (Diff) invariance, i.e. invariance under arbitrary coordinate transformations,  
is the fundamental symmetry of General Relativity \cite{Weinberg:1972kfs}. Very much as for other gauge symmetries, it determines how matter fields couple to gravity and what is the number of propagating degrees of freedom present in the theory. However, despite its central role in the construction of Einstein's theory of  gravity, it has been long known \cite{vanderBij:1981ym,Alvarez:2006uu} that the minimal symmetry
required by unitarity in theories with a massless spin-2 field is not the full Diff group, but the 
subgroup of transverse diffeomorphisms (TDiff), characterized by coordinate transformation which preserve the metric determinant, i.e. 
\begin{align}
\hat x^\mu=x^\mu+\xi^\mu(x)\label{cotr}
\end{align}
such that infinitesimally we have
\begin{align}
\partial_\mu \xi^\mu(x)=0.
\end{align}
As a matter of fact, consistent theories of gravity invariant under transverse diffeomorphisms date back to Einstein himself who proposed the so called unimodular gravity theory \cite{Einstein}
in which the determinant of the metric tensor is restricted 
to be a non-dynamical field. The corresponding field equations are the
trace-free Einstein's equations \cite{Unruh:1988in,Carballo-Rubio:2022ofy} in which vacuum energy, or any cosmological constant type contribution in the energy-momentum tensor, does not gravitate, thus providing a neat solution of the vacuum-energy problem \cite{Ellis:2010uc,Jirousek:2023gzr}. 
In addition, if the energy-momentum tensor of the matter fields
is conserved, it can be seen that the equations of motion of
unimodular gravity are just Einstein's equations supplemented with a cosmological constant term
that appears as an integration constant. Thus, the theory actually propagates the same number of local degrees of freedom as General Relativity plus a global one which is  canonically conjugate 
to the cosmological constant \cite{Henneaux:1989zc,Kuchar:1991xd}.
This connection between broken diffeomorphisms and the cosmological constant  has triggered the interest in potential descriptions of the cosmological dark sector  within this framework \cite{Pirogov:2005,Pirogov:2011,Maroto,Bello-Morales,Jaramillo-Garrido,Alonso-Lopez,Tessainer:2024ewm}. 

The purpose of this work is two-fold. On the one hand, we will analyze in Section II the limitations that Diff invariance sets in a field-theory description of the cosmological dark sector. On the other hand, in Section III, we show that these limitations can be easily avoided in the framework of TDiff theories and present a simple model with a single scalar field which is identical to $\Lambda$CDM in any background geometry. 

\section{Dark sector in a Diff invariant theory}
Diff invariance sets strong limitations for a field-theory description of the dark sector of cosmology.  Indeed, the standard $\Lambda$CDM dark sector can be described as a perfect fluid with an
evolving equation of state that in a Robertson-Walker background reads
\begin{align}
w_{\text{DS}}= \frac{p_c+p_\Lambda}{\rho_c+\rho_\Lambda} =\frac{-1}{1+\frac{\Omega_c}{\Omega_\Lambda}a^{-3}}, \label{wLCDM}
\end{align}
where $p_c$ ($p_\Lambda$) and $\rho_c$ ($\rho_\Lambda$) are the cold dark matter (cosmological constant) pressure and energy density, $\Omega_c$ ($\Omega_\Lambda$) is the corresponding density parameter and $a$ is the scale factor. 
Since a pure cosmological constant does not support pressure perturbations and the pressure of cold dark matter is negligible,  the total pressure perturbation of the fluid vanishes 
$\delta p_{\text{DS}}=0$.  Accordingly the corresponding speed of sound also vanishes $c_s^2=0$. 
However, a fluid with these properties cannot be described with a simple  Diff-invariant scalar field theory. Indeed, let us consider the most 
general Diff-invariant action for a scalar field minimally coupled to gravity \cite{Armendariz-Picon:2000nqq,Armendariz-Picon:2000ulo}
\begin{align}
S[g_{\mu\nu},\phi]&=\int d^4x\, \sqrt{g}\,{\cal L}(\phi,X)\label{Diff_action}
\end{align}
with $g=\vert \det g_{\mu\nu} \vert $ and 
 \begin{align}
X=\frac{1}{2}g^{\mu \nu}\partial_\mu\phi\,\partial_\nu \phi .
\label{X}
\end{align}
For $X>0$, the corresponding energy-momentum tensor takes the perfect fluid form
\begin{equation}\label{EMT}
    T_{\mu\nu} = (\rho+p)u_\mu u_\nu -pg_{\mu\nu}
\end{equation}
where the fluid velocity is
\begin{equation}\label{eq: velocity}
     u_\mu = \frac{\partial_\mu \phi}{\sqrt{2X}}
\end{equation}
and the energy density and pressure read
\begin{align}
\rho_\phi&=2X\partial_X {{\cal L}(\phi,X)}- {\cal L}(\phi,X)\\
p_\phi&=  {\cal L}(\phi,X)
\end{align}
so that the corresponding equation of state and speed of sound are
\cite{Armendariz-Picon:2000nqq,Armendariz-Picon:2000ulo}
\begin{align}
w_\phi&=\frac{{\cal L}(\phi,X)}{2X\partial_X {{\cal L}(\phi,X)}- {\cal L}(\phi,X)}\\
c_{s,\phi}^2&=\frac{\partial_X {{\cal L}(\phi,X)}}{\partial_X {{\cal L}(\phi,X)}+2X\partial_X^2 {{\cal L}(\phi,X)}}.
\end{align}
If we impose $c_{s,\phi}^2=0$ then $\partial_X {{\cal L}(\phi,X)}=0$ and consequently $w_\phi=-1$
which means that the equation of state parameter of the scalar field cannot mimic the 
one for $\Lambda$CDM in \eqref{wLCDM}.  Moreover, for this type of theories, in general it is not possible to guarantee that $X$ remains positive everywhere so that the perfect-fluid description can be preserved \cite{Diez-Tejedor:2006ybl}. 

Despite the previous limitations, several models have been proposed for a unified description of the dark sector with a single scalar field in the framework of k-essence 
theories \cite{Scherrer:2004au,Giannakis:2005kr,Bertacca:2007ux}.  However, such models exhibit certain deviations with respect to $\Lambda$CDM. 
The perfect mimicry was achieved in \cite{Lim:2010yk} with a theory comprising two
scalar fields, one of which is a Lagrange multiplier that enforces the energy flow along timelike geodesics. 
\section{Dark sector in a TDiff invariant theory}
In this work, we show that the mentioned obstacles can be easily overcome, even for single-field models,  if we allow for a TDiff symmetry rather than the full Diff invariance in the scalar coupling to gravity. In fact, we will show that when the full Diff symmetry is restored in these models, the extra
Lagrange multiplier fields introduced in \cite{Lim:2010yk,Sebastiani:2016ras} appear in a natural way. 

Let us then consider the following action for a single scalar field with 
canonical kinetic term \cite{Maroto,Jaramillo-Garrido}
\begin{align}
S[g_{\mu\nu},\phi]&=\int d^4x\, f(g)X\label{TDiff_action}
\end{align}
where $f(g)$ is a given positive function of the metric determinant. For $f(g)=\sqrt{g}$ we recover the scalar field action with purely canonical kinetic term, invariant under full diffeomorphisms. However for arbitrary $f(g)$, this action breaks the Diff symmetry down to the subgroup of transverse diffeomorphisms. Even though it is possible to work with the previous action, calculations are eased if we restore full Diff  symmetry  by introducing an additional (vector) Stueckelberg field  $A^\mu$ 
 \cite{Henneaux:1989zc,Jaramillo-Garrido:2024tdv}    as
 \begin{align}
S[g_{\mu\nu},\phi, A^\mu]&= \int d^4x \sqrt{g}\, H_K(Y) X\label{cov_action}
\end{align}
 where $Y=\nabla_\mu A^ \mu$ and 
 \begin{align}
 H_K(Y)=Yf(Y^{-2})
 \end{align}
 so that the TDiff action \eqref{TDiff_action} is recovered in the TDiff coordinate
 frame in which
 \begin{align}
 Y=\frac{1}{\sqrt{g}}. \label{TDiff_frame}
 \end{align}
Notice that in this notation, the Diff case $f(g)=\sqrt{g}$ corresponds
to the function $H_K(Y)=1$. 

Variations of the previous action with respect to the scalar field $\phi$ yield
\begin{align}
\nabla_\nu(H_K(Y)g^{\mu\nu}\partial_\mu\phi)=0. \label{scalarcov}
\end{align}
On the other hand,  variations with respect to the vector field $A^\alpha$ lead to
\begin{align}
\partial_\alpha\left(H_K'(Y)X\right)=0\label{Yeq}
\end{align}
where a prime denotes derivative with respect to its argument, so that 
\begin{align}
H_K'(Y)X=\lambda_0 \label{constraint}
\end{align}
with $\lambda_0$ a constant. 
Finally, variations with respect to the metric tensor allow us to obtain the corresponding 
 energy-momentum tensor
\begin{align}
T_{\mu\nu}&=H_K(Y)\partial_\mu\phi\partial_\nu\phi-(H_K(Y)-YH_K'(Y))X g_{\mu\nu}
\end{align}
where we have used \eqref{Yeq}. In the case $X>0$, this can be written as the energy-momentum tensor of a perfect fluid \eqref{EMT} where the fluid velocity is
\eqref{eq: velocity} and the energy density and pressure read \cite{Jaramillo-Garrido:2024tdv}
\begin{subequations}\label{eq: covar energy density and pressure}
    \begin{align}
        \rho &= (H_K(Y)+YH_K'(Y))X \,, \label{subeq: rho covar}\\[5pt]
        p &= (H_K(Y)-YH_K'(Y))X \,. \label{subeq: p covar}
    \end{align}
\end{subequations}
Because of the constant of motion in \eqref{constraint}, energy density and pressure depend
on a single field $Y$ and the fluid is adiabatic.  As a matter of fact, the corresponding equation of state and speed of sound read \cite{Jaramillo-Garrido:2024tdv}
\begin{align}
w&=\frac{H_K(Y)-YH_K'(Y)}{H_K(Y)+YH_K'(Y)}\\
c_s^2&=\frac{H_K(Y) H_K''(Y)}{H_K(Y) H_K''(Y)-2H_K'^2(Y)}.
\end{align}
On the other hand, combining \eqref{scalarcov} and \eqref{Yeq} it is possible to obtain 
an equation for the evolution of the $Y$ field as \cite{Jaramillo-Garrido:2024tdv}
\begin{align}
u^\mu\partial_\mu \ln (X^{1/2}H_K(Y)\delta V)=0 \label{YV}
\end{align}
where $\delta V$ is the cross-sectional volume of a congruence which is related 
to the expansion $\theta$ through
\begin{align}
 \theta= \nabla_\mu u^\mu= u^\mu \partial_\mu \ln \delta V.
 \end{align} 

Notice that unlike the Diff case, imposing $c^2_s=0$ does not force $w=-1$. The condition $c_s^2=0$ implies $H_K''(Y)=0$, i.e.
\begin{align}
H_K(Y)=\alpha +\beta Y \label{linear}
\end{align}
with $\alpha$ and $\beta$ constants, so that the corresponding equation of state reads
\begin{align}
w=\frac{\alpha}{\alpha +2\beta Y}. \label{wab}
\end{align}
On the other hand, the constant of motion \eqref{constraint}, implies in this case
\begin{align}
X=\frac{\lambda_0}{\beta} \label{X}
\end{align}
so that, provided $\beta\neq 0$, $X$ is constant and the fluid velocity can be timelike everywhere. Therefore in this TDiff-invariant case it is possible to preserve the perfect fluid description. 

Without loss of generality we can set $H_K(Y=1)=1$, so that $\beta=1-\alpha$.
Notice that since $X$ is constant, \eqref{YV} implies that along the fluid trajectories 
$H_K(Y)\delta V$ remains constant in any background geometry. Accordingly, 
since the cross-sectional volume is in the interval $\delta V\in [0,\infty)$, for positive constant, the  $Y$ field must be in the range $Y\in [\frac{\alpha}{\alpha-1},\infty)$, which translates
into $\sqrt{g}\in (0,\frac{\alpha-1}{\alpha}]$ in the TDiff coordinate frame,  which is a positive interval for $\alpha\leq 0$ or 
$\alpha>1$.  In particular, for $\alpha\leq 0$ the TDiff volume function corresponding to the linear expression \eqref{linear} reads 
\begin{align}
f(g)=(1-\alpha)+\alpha\sqrt{g} \label{f(g)}
\end{align}
which is non-negative everywhere and the same happens with the energy density
\begin{align}
\rho=(\alpha+2(1-\alpha)Y)X
\end{align}
provided the fluid velocity is timelike.

In a Robertson-Walker background 
\begin{align}
ds^2=dt^2-a^2(t)d\vec x^2
\end{align}
the cross-sectional volume satisfies $\delta V\propto a^3$, so that \eqref{YV}
implies $H_K(Y)=a^{-3}$, where we have set for simplicity $Y(a=1)=1$. 
In other words, from \eqref{linear} we get
\begin{align}
Y(a)=-\frac{\alpha}{1-\alpha}+\frac{a^{-3}}{1-\alpha}
\end{align}
so that substituting in \eqref{wab} we recover the $\Lambda$CDM equation of state
\eqref{wLCDM} with
\begin{align}
\alpha=-2\frac{\Omega_\Lambda}{\Omega_c}.
\end{align}
 Thus, we conclude that a purely kinetic scalar field with canonical kinetic term and a coupling to gravity given by \eqref{f(g)}  mimics the dynamics of $\Lambda$CDM at the background and perturbation levels.
 
In fact, we can reach the same conclusion directly from the energy-momentum tensor, 
which reads
 \begin{align}
T_{\mu\nu}&=(\alpha+(1-\alpha)Y)\partial_\mu\phi\partial_\nu\phi+\alpha X g_{\mu\nu}
\end{align}
and can be written as the sum of two conserved energy-momentum tensors
\begin{align}
T_{\mu\nu}=T_{\mu\nu}^{c}+ T_{\mu\nu}^\Lambda
\end{align}
with
\begin{align}
 T_{\mu\nu}^\Lambda=\alpha X g_{\mu\nu}.
\end{align}
Notice that in this case  the constraint equation \eqref{constraint}
implies $X=$ const. so that   $T_{\mu\nu}^\Lambda$ is independently conserved, with 
\begin{align}
\rho_\Lambda=-p_\Lambda=\alpha X \label{lambda}.
\end{align}
On the other hand, 
\begin{align}
 T_{\mu\nu}^{c}=2X(\alpha+(1-\alpha)Y)u_\mu u_\nu
\end{align}
also takes the perfect fluid form with
\begin{align}
\rho_{c}&=2X (\alpha+(1-\alpha)Y)\\
p_c&=0.
\end{align}
Thus the total energy-momentum tensor is the sum of a cosmological constant contribution and 
an irrotational dust fluid in any background geometry. 

It is also interesting to analyze the different relevant limits of the coupling function  \eqref{f(g)}. Thus, for $\alpha=1$ we recover the Diff invariant 
case, in which the scalar field behaves as a stiff fluid 
with $w=1$. For $\alpha=0$, i.e. $f(g)=1$, the model describes pure cold dark matter \cite{Maroto,Jaramillo-Garrido}.  On the other hand, a pure cosmological constant with no dark matter can only be obtained as a limit when $\alpha\rightarrow -\infty$.

In order to identify the number of local degrees of freedom present in the model, we can rewrite the action as \cite{Bello-Morales:2024vqk}
\begin{align}
S&=\int d^4x \sqrt{g} (\alpha+\beta Y)X \nonumber \\
&= \int d^4x \sqrt{g} \left[(\alpha+\beta \psi)X+\lambda_0(\nabla_\mu A^\mu-\psi)\right]
\end{align}
where we have introduced the new scalar field $\psi$ through the Lagrange multiplier $\lambda_0$.  The equation of motion for $A^\mu$ then imposes $\lambda_0=$ const. , 
whereas the equation of motion for the $\psi$ field, which is also a Lagrange multiplier, 
implies $\beta X=\lambda_0$ in agreement with \eqref{X}. Since $\lambda_0$ is a constant then the $\nabla_\mu A^\mu$ term is just a surface term and we can write
\begin{align}
S&= \int d^4x \sqrt{g} \left[\alpha X+\psi(\beta X -\lambda_0)\right].
\end{align}
Thus, we recover the action in \cite{Lim:2010yk} in which we only have one local degree  of freedom 
corresponding to the field $\phi$, whereas $\psi$ is just a Lagrange multiplier. Notice that unlike \cite{Lim:2010yk}
in which $\lambda_0$ appeared as a constant parameter of the action, here it 
is a global degree of freedom which is fixed by the initial conditions in the fields, very much as the cosmological constant in unimodular gravity.    

Since the speed of sound identically vanishes, there are no 
wave-like dynamical degrees of freedom at the perturbative level in the model with $X>0$ \cite{Lim:2010yk}. However, the complete equations \eqref{scalarcov} and \eqref{Yeq} support null waves with $X=0$. Indeed, it is straightforward to check that in Minkowski space-time, $\phi\propto e^{ikx}$ is a solution with $k^2=0$ for 
$Y=1$ with $\lambda_0=0$, although the non-linear constraint prevents the linear superposition of solutions. 

Given the similarities with unimodular gravity, it is also interesting to  determine whether in this case it is also possible to define a TDiff {\it time} field conjugate  to the cosmological constant \cite{Henneaux:1989zc}.  Thus, given some initial space-like hypersurface $\Sigma(t_0)$ we define the TDiff time 
associated to the $\Sigma(t)$ hypersurface by
\begin{align}
T(t)&=\int_{\cal R} d^4x \sqrt{g}\, \nabla_\mu A^\mu= \int_{\cal R} d^4x \,\partial_\mu (\sqrt{g}A^\mu)\nonumber \\
&=\int_{\Sigma(t)} d^3x\, {\cal A}^0- \int_{\Sigma(t_0)} d^3x\, {\cal A}^0
\end{align}
where ${\cal R}$ denotes the spacetime region between $\Sigma(t)$ and $\Sigma(t_0)$
 and we have defined the vector density ${\cal A}^\mu=\sqrt{g} A^\mu$. We observe 
that in the TDiff frame given by \eqref{TDiff_frame}, the TDiff time is thus understood as the spacetime volume between the hypersurfaces. In addition, 
 from  the action \eqref{cov_action}, we can see that the conjugate momentum of the {\it time} field ${\cal A}^0$, reads
 \begin{align}
 \pi_{{\cal A}^0}=\frac{\partial {\cal L}}{\partial (\partial_0{\cal A}^0)}= (1-\alpha)X
 \end{align}
which is also proportional in this case to the cosmological constant in \eqref{lambda}.

\section{Conclusions and prospects}
In this work we have shown that simply by replacing
$\sqrt{g}\rightarrow (1-\alpha)+\alpha\sqrt{g}$
in the integration measure of the action,  a single scalar field with canonical kinetic term becomes a perfect mimicker of $\Lambda$CDM in any background geometry. The $\alpha$ parameter sets the value of the corresponding $\Omega_c$ parameter in $\Lambda$CDM. The resulting action breaks Diff symmetry down to the subgroup of transverse diffeomorphisms. This result suggests that certain classes of TDiff scalar fields with canonical kinetic terms could provide simple unified descriptions of the dark sector in cosmology, thus overcoming the traditional limitations of single-field Diff scalars. Moreover, this type of theories 
could evade the stringent limits imposed by 
large-scale structure observations on the dark sector 
speed of sound \cite{Sandvik:2002jz}. In particular, for coupling functions different from \eqref{f(g)},  deviations from $\Lambda$CDM compatible with current observations are also possible \cite{Cruz2025}, thus offering a new framework for the description of dynamical dark energy.

\acknowledgements{We would like to thank Jose Beltr\'an Jim\'enez, Dar\'{\i}o Jaramillo-Garrido, Prado Mart\'{\i}n-Moruno and Diego Tessainer for useful comments and suggestions. The research of JdCP was financially supported by the project "Plan Complementario de I+D+i en el \'area de Astrof{\'\i}sica" funded by the European Union within the framework of the Recovery, Transformation and Resilience Plan - NextGenerationEU and by the Regional Government of Andaluc{\'i}a (Reference AST22\_00001). This work has been supported by the MICIN
(Spain) Project No. PID2022-138263NB-I00 funded by
MICIU/AEI/10.13039/501100011033 and by ERDF/EU}

\bibliographystyle{apsrev4-1}
\bibliography{bibliography}

\end{document}